\documentclass[aps,showpacs,prc,draft,12pt]{revtex4}

\usepackage{epsfig,bm}

\newcommand{\ba}{\begin{eqnarray}}
\newcommand{\ea}{\end{eqnarray}}
\newcommand{\bmath}{\begin{mathletters}}
\newcommand{\emath}{\end{mathletters}}
\newcommand{\ban}{\begin{eqnarray*}}
\newcommand{\ean}{\end{eqnarray*}}

\begin{document}

\title{X(5) Critical-Point Structure in a Finite System}

\author{A. Leviatan}

\affiliation{
Racah Institute of Physics, The Hebrew University,
Jerusalem 91904, Israel}

\date{\today}

\begin{abstract}
$X(5)$ is a paradigm for the structure at the critical point of a 
particular first-order phase transition for which the intrinsic energy 
surface has two degenerate minima separated by a low barrier. 
For a finite system, we show that the dynamics at such a critical point 
can be described by an effective deformation 
determined by minimizing the energy surface after projection onto 
angular momentum zero, and combined with two-level mixing.
Wave functions of a particular analytic form are used to 
derive estimates for energies and quadrupole rates at the 
critical point. 
\end{abstract}


\maketitle

\newpage

Quantum phase transitions, occurring at zero temperature as a function of a 
coupling constant, have become a topic of great interest in different 
branches of physics~\cite{iac00,iac01,voj03}.
Often this type of phase transition involves a 
structural change between different shapes or geometric configurations.
Advanced experiments have identified such quantum shape-phase transitions 
in a variety of mesoscopic systems, with a finite number of constituents, 
{\it e.g.}, nuclei, molecules and atomic clusters. 
A key issue is to understand the modifications brought in by the 
finiteness of these systems near criticality. This question can be 
conveniently addressed in a class of quantum models in which 
the Hamiltonian is expanded in elements of a Lie algebra.
Such models are widely used in the 
description of nuclei and molecules~\cite{ibm,vibron}. 
In the present work we study this question in connection with nuclei, 
exemplifying a finite system undergoing a first-order shape-phase 
transition. The tools developed are applicable to other 
mesoscopic systems described by similar models albeit with different 
spectrum generating algebras. 

Recently, it has been recognized that 
quantum shape-phase transitions are amenable to analytic descriptions 
at the critical points~\cite{iac00,iac01}. 
The importance of these analytic benchmarks of criticality 
lies in the fact that they provide a classification of states and analytic 
expressions for observables in regions 
where the structure changes most rapidly.
For nuclei these benchmarks 
were obtained in the geometric framework 
of a Bohr Hamiltonian for macroscopic quadrupole shapes.
In particular, the E(5) \cite{iac00} (X(5) \cite{iac01}) benchmark 
is applicable to a second- (first-) order phase transition between
spherical and deformed $\gamma$-unstable (axially-symmetric) nuclei. 
In the present work we focus on the X(5) benchmark 
for which an empirical example has been found in 
$^{152}$Sm, $^{150}$Nd, 
$^{156}$Dy and $^{154}$Gd~\cite{casten01,krucken02,caprio02a,tonev04}.
The role of a finite number of nucleons can be addressed in 
the algebraic framework of the interacting boson model 
(IBM)~\cite{ibm} which describes low-lying quadrupole collective 
states in nuclei in terms of a system of $N$ monopole ($s$) and
quadrupole ($d$) bosons representing valence nucleon pairs. 
The model has U(6) as a spectrum generating algebra and its three 
dynamical symmetry limits: U(5), SU(3), and O(6), 
describe the dynamics of stable nuclear 
shapes: spherical, axially-deformed, and $\gamma$-unstable deformed. 
Phase transitions for finite N are studied 
by an IBM Hamiltonian involving terms from different dynamical 
symmetry chains~\cite{diep80}. 
An important conclusion from many such 
studies~\cite{izc98,ckz99,zam02,levgin03,arias03,iaczam04,rowe04} 
is that although in finite systems the discontinuities at the critical 
point are smoothed out, features of the phase transition persist even 
at moderate values of $N$ ($N=10$ for $^{152}$Sm).

The original formulation 
of the E(5) and X(5) models~\cite{iac00,iac01} employed an 
infinite square-well potential in the $\beta$ variable of the Bohr 
Hamiltonian. This is an adequate approximation for an E(5)-like 
second-order phase transition, where the relevant potential 
is $\gamma$-indpendent and flat-bottomed. 
This behaviour persists in the finite-N 
energy surface, obtained by the method of 
coherent states~\cite{diep80,gino80}. 
In this case, the large fluctuations in $\beta$ 
can be taken into account by means of an effective $\beta$-deformation 
determined by minimizing the energy surface after projection onto 
the appropriate [$O(5)$] symmetry~\cite{levgin03}. 
The structure of X(5)-like first-order phase transition is more complex. 
The geometric $X(5)$ solution~\cite{iac01} 
assumes the decoupling of the $\beta$ and 
$\gamma$ degrees of freedom. The corresponding 
finite-N energy surface 
displays two degenerate minima separated by a low barrier. 
Analysis of the IBM wave functions near the critical-point 
show clear evidence for 
phase coexistence and level crossing~\cite{izc98,zam02,arias03}. 
This indicates that although the barrier 
between the two minima is small, its effect 
cannot be completely ignored. 
In the present work we examine the conditions for and properties of 
$X(5)$ critical-point structure in a finite system. 

Focusing on the dynamics of the $\beta$ degree of freedom, 
in the geometric approach, the X(5) eigenfunctions~\cite{iac01} 
are proportional to Bessel functions, 
and the spectrum consists of families of states, $L^{+}_{\xi}$, 
labeled by $\xi=1,2,\ldots$, with angular momentum 
$L=0,2,4,\ldots$ and projection $K=0$ along the symmetry-axis. 
The X(5) benchmark leads to analytic parameter-free
predictions for energy ratios and $B(E2)$ ratios 
which, as seen in Table I, are in-between the values expected of a 
spherical vibrator [$U(5)$] and an axially-deformed rotor [$SU(3)$].

In the algebraic approach, the $U(5)$-$SU(3)$ transition 
is modeled by the Hamiltonian
\ba
H &=& \epsilon\,\hat{n}_d -\kappa\, Q\cdot Q ~.
\label{hamilt}
\ea
Here $\hat{n}_d$ is the 
$d$-boson number operator, $Q$ is the quadrupole generator of $SU(3)$ 
and the dot implies a scalar product.
In the $U(5)$ limit ($\kappa=0$), 
the spectrum of $H$ is harmonic, 
and the eigenstates are 
classified according to the chain
$U(6)\supset U(5)\supset O(5) \supset O(3)$ with quantum numbers
$\vert\,N,n_d,\tau,L\rangle$ (for $\tau\geq 6$ 
an additional multiplicity index may be required for complete classification).
In the $SU(3)$ limit ($\epsilon=0$), the Hamiltonian is related 
to the Casimir operator of $SU(3)$ and the spectrum is indicated in 
the caption of Table I. 
The eigenstates are classified according to the chain
$U(6)\supset SU(3)\supset O(3)$ with quantum numbers
$\vert\,N,(\lambda,\mu),K,L\rangle$. 
A geometric visualization is obtained by 
an intrinsic energy surface defined by 
the expectation value of the Hamiltonian in the coherent 
state~\cite{diep80,gino80}
\ba
\vert\,\beta,\gamma ; N \rangle &=&
(N!)^{-1/2}(b^{\dagger}_{c})^N\,\vert 0\,\rangle ~,
\label{cond}
\ea
where $b^{\dagger}_{c} = (1+\beta^2)^{-1/2}[\beta\cos\gamma\,
d^{\dagger}_{0} + \beta\sin{\gamma}\,
( d^{\dagger}_{2} + d^{\dagger}_{-2})/\sqrt{2} + s^{\dagger}]$ with 
$\beta\geq 0$ and $0\leq\gamma\leq\pi/3$. 

The IBM Hamiltonian at the critical point of the
$U(5)$-$SU(3)$ phase transition corresponds to the following 
choice of parameters~\cite{diep80} in the Hamiltonian of 
Eq.~(\ref{hamilt})
\ba
\epsilon &=& \frac{9}{4}\kappa (2N-3) ~.\;
\label{hcri}
\ea
Its intrinsic energy surface given by
\ba
E(\beta,\gamma) &=& 
-5\kappa N +
\frac{\kappa\,N(N-1)\,\beta^2}{2(1+\beta^2)^{2}}
\left ( 1 -4\sqrt{2}\beta\cos3\gamma + 8\beta^2 \right ), 
\label{enesurf}
\ea 
has the typical form of a Landau potential for a first-order phase 
transition, with two degenerate minima, at $\beta=0$ and at 
$(\beta=\frac{1}{2\sqrt{2}},\gamma=0)$. 
As shown in Fig.~(1),   
the barrier separating the spherical and prolate-deformed minima is 
extremely small and the resulting surface, 
$E(\beta)\equiv E(\beta,\gamma=0)$, is rather flat. 
This behaviour motivated the use of a square-well potential in the 
X(5) model~\cite{iac01}. 
Guided by the experience gained with finite-N flat-bottomed potentials 
in second-order phase transitions~\cite{levgin03}, 
we are led to consider 
states, $\vert\, \beta; N, L\rangle$, of good $O(3)$ symmetry $L$ 
projected from the intrinsic state $\vert \beta,\gamma=0;N\rangle$ of 
Eq.~(\ref{cond}), with an effective $\beta$-deformation yet to be 
determined. In the $U(5)$ basis these $L$-projected states are 
\ba
\vert\, \beta; N, L\rangle &=&
\sum_{n_d,\tau}\frac{1}{2}
\left [1 + (-1)^{n_d-\tau}\right ]\, \xi_{n_d,\tau}^{(N,L)}
\vert\, N,n_d,\tau, L\rangle ~,
\label{projndt}
\ea
where $L$ is even, $\tau$ takes the values compatible with the 
$O(5)\supset O(3)$ reduction and the $n_d$ summation covers the range 
$\tau\leq n_d\leq N$.
The coefficients $\xi_{n_d,\tau}^{(N,L)}$ are of the form 
\ba
\xi_{n_d,\tau}^{(N,L)} &=& 
\left [\Gamma_{N}^{(L)}(\beta)\right ]^{-1/2} 
f_{\tau}^{(L)}\frac{\beta^{n_d}}
{\left [
(N-n_d)!(n_d-\tau)!!(n_d +\tau +3)!!\right ]^{1/2}}
\label{xindt}
\ea
where $\Gamma_{N}^{(L)}(\beta)$ is a normalization factor.
In some cases analytic expressions of these coefficients can be derived. 
Specifically, for $L=0$: 
$f_{\tau}^{(0)}= (-1)^{\tau}\sqrt{2\tau+3}$ with $\tau=0,3,6,\ldots$ 
and for $L=2$: 
$f_{\tau}^{(2)}= (-1)^{\tau+1}\sqrt{\tau+2}$ 
($\,= (-1)^{\tau+1}\sqrt{\tau+1}\,$)
with $\tau=1,4,7,\ldots$ 
(with $\tau=2,5,8,\ldots$). 
In general, $\xi_{n_d,\tau}^{(N,L)}$ can be obtained by numerical 
diagonalization of Hamiltonians which have the condensate (\ref{cond}) 
as an exact eigenstate~\cite{kirlev85}. 
The states $\vert\, \beta; N, L\rangle$ interpolate between the 
$U(5)$ spherical ground state, $\vert s^N\rangle$, with $n_d=\tau=L=0$, 
at $\beta=0$, 
and the $SU(3)$ deformed ground band with $(\lambda,\mu)=(2N,0)$, 
at $\beta=\sqrt{2}$. 
For arbitrary $\beta$ the matrix elements of the Hamiltonian 
in these states define an $L$-projected energy surface~\cite{Hag03}, 
$E^{(N)}_{L}(\beta) 
= \langle \beta; N, L\vert H \vert \beta; N, L\rangle $,
which can be evaluated in closed form
\ba
E^{(N)}_{L}(\beta) &=& 
\epsilon\left[\, N - S^{(N)}_{1,L}\,\right ]
\nonumber\\
&&\;
+ \frac{1}{2}\kappa\,\left [\, 
(\beta^2 - 2)^2\, S^{(N)}_{2,L}
+ 2 (\beta - \sqrt{2})^2\, \Sigma^{(N)}_{2,L} + 
\frac{3}{4}L(L+1) - 2N(2N+3)\, \right ] ~.
\quad
\label{egL}
\ea
The quantities
$S^{(N)}_{1,L}$, $S^{(N)}_{2,L}$ and $\Sigma^{(N)}_{2,L}$ are, 
respectively, the expectation values of 
$\hat{n}_s=s^{\dagger}s$, $(s^{\dagger})^2s^2$ and 
$\hat{n}_s\,\hat{n}_d$ in the states $\vert \beta;N,L\rangle$. They are 
determined by the normalization factor $\Gamma^{(L)}_{N}(\beta)$ of 
Eq.~(\ref{xindt}) 
\ba
S^{(N)}_{1,L} &=& \frac{\Gamma^{(L)}_{N-1}(\beta)}
{\Gamma^{(L)}_{N}(\beta)} ~,
\ea
with $S^{(N)}_{2,L} = S^{(N)}_{1,L}S^{(N-1)}_{1,L}$ and
$\Sigma^{(N)}_{2,L} =
(N-1)S^{(N)}_{1,L} - S^{(N)}_{2,L}$.

As noted in~\cite{jolie99} and shown in Fig.~(1b), 
the $L=0$ projected energy surface, $E^{(N)}_{L=0}(\beta)$, 
no-longer exhibits the double 
minima structure observed in the (unprojected) intrinsic energy 
surface. Instead, there is a minimum at $\beta>0$, 
a maximum at $\beta=0$, and a saddle point in the $\gamma$ 
direction at $\beta<0$. 
$E^{(N)}_{L=2}(\beta)$ 
resembles the potential used in the geometric collective model 
calculation of 
$^{152}$Sm~\cite{cap99} with 
a minimum at a larger value of $\beta>0$, 
and a flat shoulder near $\beta=0$. 
The different behaviour of 
$E^{(N)}_{L=2}(\beta)$ and $E^{(N)}_{L=0}(\beta)$ 
can be attributed to the fact that, as shown 
in Fig.~(1a), the $L=2$ state is 
well above the barrier 
and hence experiences essentially a flat-bottomed potential.
In contrast, 
the two minima in the intrinsic energy surface support two 
coexisting $L=0$ states which, in view of the low barrier, 
are subject to considerable mixing. 
The mixing between the spherical and deformed $L=0$ states
\ba
\vert\phi_1\rangle &\equiv& \vert s^N\rangle ~,
\nonumber\\
\vert \phi_2\rangle &\equiv& \vert \beta;N,L=0\rangle ~, 
\ea
can be studied by means of a $2\times 2$ potential energy matrix, 
$m_{ij}=\langle\phi_i\vert H \vert \phi_{j}\rangle$, given by 
\ba
m_{11} &=& -5\kappa\,N\;\; , \;\; 
m_{12} = -\kappa N\left [\beta^2(N-1) + 5 \right ]r_{12} ~,
\nonumber\\
m_{22} &=& E^{(N)}_{L=0}(\beta)\;\; , \;\;
r_{12} = \left [ N!\,\Gamma^{(L=0)}_{N}(\beta)\,\right ] ^{-1/2} ~,
\ea 
where $r_{12}=\langle\phi_1\vert \phi_2\rangle$ is the overlap. 
The eigenpotentials are obtained by diagonalization and the equilibrium 
$\beta$-deformation is determined from the minimum of the lowest 
eigenpotential. This procedure is in the spirit of the matrix coherent-state 
approach proposed in~\cite{frank04} to describe the geometry of 
configuration mixing in nuclei. An important difference being that 
instead of using coherent states with different values of N, in the present 
study we consider states with good angular momentum, $L=0$, and fixed N. 
A slight complication arises 
from the fact the two states $\vert\phi_1\rangle$ and $\vert\phi_2\rangle$ 
are not orthogonal. 
This can be taken into account by 
transforming into an orthonormal basis of $L=0$ states 
\ba
\vert \Psi_1\rangle &=& \vert\phi_1\rangle ~,
\nonumber\\
\vert \Psi_2\rangle &=& 
(1-r_{12}^2)^{-1/2}\,
\left(\,\vert\phi_2\rangle - r_{12}\,\vert\phi_1\rangle\, \right ) ~.
\ea 
The matrix elements 
$K_{ij} = \langle \Psi_{i}\vert H \vert \Psi_{j} \rangle$ 
are then given by 
\ba
K_{11} &=& m_{11} \; , \; 
K_{12} = (1-r_{12}^2)^{-1/2}
\left ( \, m_{12} - r_{12}\, m_{11} \, \right )
\nonumber\\
K_{22} &=& (1-r_{12}^2)^{-1}
\left ( \, m_{22} - 2r_{12}\,m_{12} + r_{12}^2 \, m_{11}\, \right )~.
\ea
The eigenvalues define the eigenpotentials 
\ba
E^{(\pm)}_{L=0}(\beta) &=& 
\left ( K_{11} + K_{22} \pm \Delta \right )/2 ~,
\nonumber\\
\Delta &=& \sqrt{(K_{22}-K_{11})^2 + 4K_{12}^2} ~,
\label{evL0}
\ea
and the corresponding eigenvectors are given by
\ba
\vert \Phi^{(-)}_{L=0}\rangle &=& 
\sin\theta \,\vert \Psi_1\rangle
+ \cos\theta \,\vert \Psi_2\rangle ~,
\nonumber\\
\vert \Phi^{(+)}_{L=0}\rangle &=& 
\cos\theta \,\vert \Psi_1\rangle
- \sin\theta \,\vert \Psi_2\rangle ~,
\nonumber\\
\tan\theta &=& \frac{2K_{12}}{K_{22}-K_{11} - \Delta}~.
\label{efL0}
\ea
The lowest eigenpotential, $E^{(-)}_{L=0}(\beta)$, is shown in Fig.~(1b) 
for $N=10$. It has a minimum at a value of $\beta=0.591$, 
larger than the minimum 
($\beta= 0.528$) of the (unmixed) $L=0$ projected energy surface
and in close proximity to the minimum ($\beta=0.619$) of the 
$L=2$ projected energy surface. We now identify the members of the 
ground band ($\xi=1$) as 
$\vert 0^{+}_1 \rangle = \vert \Phi^{(-)}_{L=0}\rangle$, Eq.~(\ref{efL0}), 
with energy $E^{(-)}_{L=0}(\beta)$, Eq.~(\ref{evL0}), 
and for $L>0$ even, $\vert L^{+}_{1} \rangle = \vert \beta; N, L \rangle$, 
Eq.~(\ref{projndt}), with energy $E^{(N)}_{L}(\beta)$, Eq.~(\ref{egL}). 
The bandhead of the 
first excited band ($\xi=2$) corresponds to 
$\vert 0^{+}_2 \rangle = \vert \Phi^{(+)}_{L=0}\rangle$, Eq.~(\ref{efL0}), 
with energy $E^{(+)}_{L=0}(\beta)$, Eq.~(\ref{evL0}). 
The value of $\beta$ used in the indicated wave functions and energies 
is chosen at the global minimum of $E^{(-)}_{L=0}(\beta)$. 

Having at hand explicit expressions for the wave functions, we can evaluate 
$E2$ matrix elements. 
To conform with the geometric $X(5)$ model~\cite{iac01}, 
we employ an IBM quadrupole operator, 
$T(E2) = d^{\dagger}s + s^{\dagger}\tilde{d}$, linear in the deformation.
For transitions involving the $L=0^{+}_1,\, 2^{+}_1$ and $0^{+}_2$ 
states, analytic expressions can be derived for $B(E2)$ values by means 
of the matrix elements
$T_1\equiv\langle \beta; N, L^{\prime}=2\vert\vert\, T(E2)\, 
\vert \vert \beta; N,L=0\rangle$ and
$T_2\equiv\langle \beta; N, L^{\prime}=2\vert\vert\, T(E2)\, 
\vert\vert s^N \rangle$, 
\ba
T_1 &=&
\frac{\beta
\left [ \Gamma^{(2)}_{N-1}(\beta) + \Gamma^{(0)}_{N-1}(\beta)\right ]}
{\left [ \Gamma^{(2)}_{N}(\beta)\,\Gamma^{(0)}_{N}(\beta)\right ]^{1/2}}~,
\nonumber\\
T_2 &=& 
\frac{\beta N}
{\left [ N!\,\Gamma^{(2)}_{N}(\beta)\, \right ]^{1/2}}~.
\label{t1t2}
\ea

To test the suggested procedure we compare in Table II the $U(5)$
decomposition of exact eigenstates obtained from numerical diagonalization
of the critical Hamiltonian, Eq.~(\ref{hcri}), for $N=10$, 
with that calculated using the $L$-projected states 
with $\beta=0.591$ [the global minimum of $E^{(-)}_{L=0}(\beta)$].
As can be seen, the latter provide an accurate approximation to 
the exact eigenstates for yrast states 
(the overlaps between calculated and exact states 
are $99.6,\,99.1,\,98.0,\,97.0,\,96.6,\,96.65\,\%$ for 
$L=0^{+}_1,\,2^{+}_1,\,4^{+}_1,\,8^{+}_1,\,10^{+}_1$, respectively). 
This agreement in the structure of wave functions is
translated also
into an agreement in energies and B(E2) values as
shown in Table I.
For the $L=0^{+}_2$ state, the agreement is fair but less precise, 
in view of the smaller overlap ($81.9\%$). 
It appears that this non-yrast state is 
affected by additional states beyond the two-state mixing considered here.
The results of Table I and II clearly demonstrate
the ability of the suggested procedure to provide faithful 
estimates to the exact finite-N calculations of the critical
IBM Hamiltonian, which
in-turn capture the essential features of the $X(5)$
critical-point structure relevant to $^{152}$Sm (which is slightly 
past the phase transition towards $SU(3)$~\cite{casten01}).

To summarize, we have examined the structure at 
the $X(5)$ critical-point of a first-order shape-phase transition 
in a finite system, by means of an 
effective $\beta$-deformation, determined by variation 
after $L$-projection and two-level mixing. 
The same procedure can be used throughout the coexistence 
region, where the two minima in the potential 
coexist but are not necessarily degenerate.
In this case, the control parameter $\kappa/\epsilon$ 
in Eq.~(\ref{hamilt}) spans the range 
between the spinodal point (where the second minimum shows up) 
and the antispinodal point (where the first minimum disappears),  
which embrace the critical point. Although we have treated explicitly 
the case of atomic nuclei described by the IBM, the same approach 
is applicable to first-order phase transitions in other mesoscopic systems,  
such as polyatomic molecules described by the algebraic vibron 
model~\cite{vibron}. This work was supported by the Israel 
Science Foundation.

\clearpage

\begin{table}
\caption{
Excitation energies (in units of $E(2^{+}_{1})=1$) 
and B(E2) values (in units of
$B(E2;\, 2^{+}_{1}\to 0^{+}_{1})=1$)
for the X(5) critical-point benchmark {\protect\cite{iac01}}, 
for several N=10 calculations, and for the experimental data of 
$^{152}$Sm {\protect\cite{casten01}}. The finite-N calculations involve 
the exact diagonalization of the critical 
IBM Hamiltonian [Eq.~(\ref{hcri})], $L$-projected states 
[Eqs.~(\ref{egL}), (\ref{evL0}) and (\ref{t1t2}) with $\beta= 0.591$], 
the $U(5)$ limit [$\epsilon\,n_d$] and the $SU(3)$ limit 
$(\kappa/2)[ -\lambda^2-\mu^2-\lambda\mu - 3\lambda -3\mu 
+ 3L(L+1)/4]$.
\normalsize}
\vskip 10pt
\begin{ruledtabular}
\begin{tabular}{lcccccc}
& X(5) & exact & $L$-projection & $U(5)$ & $SU(3)$ & $^{152}$Sm \\
&      & N=10  & N=10           & N=10   & N=10   & exp \\
\hline
$E(4^{+}_{1})$  & 2.91  & 2.43 & 2.46 & 2  & 3.33  & 3.01 \\
$E(6^{+}_{1})$  & 5.45  & 4.29 & 4.33 & 3  & 7.00  & 5.80 \\
$E(8^{+}_{1})$  & 8.51  & 6.53 & 6.56 & 4  & 12.00 & 9.24  \\
$E(10^{+}_{1})$ & 12.07 & 9.12 & 9.13 & 5  & 18.33 & 13.21 \\
$E(0^{+}_{2})$  & 5.67  & 2.64 & 3.30 & 2  & 25.33 & 5.62   \\
\hline
$4^{+}_{1}\to 2^{+}_{1}$ &
1.58 & 1.61 & 1.60 & 1.8 & 1.40 & 1.45 \\
$6^{+}_{1}\to 4^{+}_{1}$ & 
1.98 & 1.85 & 1.80 & 2.4 & 1.48 & 1.70  \\
$8^{+}_{1}\to 6^{+}_{1}$ & 
2.27 & 1.92 & 1.87 & 2.8 & 1.45 & 1.98 \\
$10^{+}_{1}\to 8^{+}_{1}$  & 
2.61 & 1.87 & 1.86 & 3.0 & 1.37 & 2.22  \\
$0^{+}_{2}\to 2^{+}_{1}$ & 
0.63 & 0.78 & 0.61 & 1.8 & 0.07 & 0.23 \\
\end{tabular}
\end{ruledtabular}
\end{table}

\clearpage

\begin{table}
\caption[]{$U(5)$ decomposition (in \%) of yrast $L^{+}_{1}$ 
states for $N=10$. 
The exact values are obtained from numerical diagonalization of the 
critical IBM Hamiltonian, Eq.~(\ref{hcri}).
The calculated values are obtained from the $L$-projected states, 
Eqs.~(\ref{xindt}) and (\ref{efL0}) with $\beta=0.591$. 
\normalsize}
\vskip 10pt
\begin{ruledtabular}
\begin{tabular}{ccccccc|ccccccc}
&\multicolumn{2}{c}{$0^{+}_{1}$}
&
\multicolumn{2}{c}{$2^{+}_{1}$}
&
\multicolumn{2}{c|}{$4^{+}_{1}$}
& &
\multicolumn{2}{c}{$6^{+}_{1}$}
& 
\multicolumn{2}{c}{$8^{+}_{1}$}
& 
\multicolumn{2}{c}{$10^{+}_{1}$}
\\
\cline{2-3}\cline{4-5}\cline{6-7}\cline{9-10}\cline{11-12}\cline{13-14}
$(n_d,\tau)$
&
exact & calc & exact & calc & exact & calc $\;$&
$(n_d,\tau)$
&
exact & calc & 
exact & calc & 
exact & calc  
\\
\hline
(0,0) & 52.3  & 52.0  &      &      &       &       &
(3,3) & 43.0  & 54.1  &      &      &       &       \\
(1,1) &       &       & 41.8 & 41.1 &       &       &
(4,4) & 23.9  & 26.4  & 46.9 & 60.1 &       &       \\
(2,0) & 31.65 & 30.15 &      &      &       &       &
(5,3) & 19.25 & 12.6  &      &      &       &       \\
(2,2) &       &       & 13.5 & 18.4 & 40.8  & 47.8  &
(5,5) & 2.3   & 2.1   & 26.7 & 26.5 & 52.2  & 66.2  \\
(3,1) &       &       & 26.5 & 25.7 &       &       &
(6,4) & 7.7   & 3.7   & 15.5 & 8.45 &       &       \\
(3,3) & 5.2   & 8.0   &      &      & 19.8  & 24.25 $\;$& 
(6,6) & 0.4   & 0.2   & 3.1  & 2.3  & 28.1  & 25.1  \\
(4,0) & 7.9   & 7.35  &      &      &       &       & 
(7,3) & 2.2   & 0.6   &      &      &       &       \\
(4,2) &       &       & 7.3  & 7.0  & 22.9  & 18.1  &       
(7,5) & 0.5   & 0.2   & 5.8  & 2.15 & 11.7  & 5.4   \\
(4,4) &       &       & 2.3  & 2.5  & 1.1   & 1.3   & 
(7,7) & 0.2   & 0.1   & 0.2  & 0.1  & 3.4   & 2.05  \\
(5,1) &       &       & 5.75 & 3.7  &       &       &
(8,4) & 0.5   & 0.1   & 1.1  & 0.2  &       &       \\
(5,3) & 1.8   & 1.9   &      &      & 8.6   & 5.6   & 
(8,6) & 0.05  & 0.0   & 0.4  & 0.1  & 3.7   & 1.1   \\
(5,5) &       &       & 0.4  & 0.4  & 1.0   & 0.7   & 
(8,8) & 0.0   & 0.0   & 0.1  & 0.0  & 0.1   & 0.0   \\
(6,0) & 0.8   & 0.5   &      &      &       &       &
(9,3) & 0.05  & 0.0   &      &      &       &      \\
(6,2) &       &       & 1.1  & 0.6  & 3.8   & 1.5   &
(9,5) & 0.0   & 0.0   & 0.2  & 0.0  & 0.4   & 0.05  \\
(6,4) &       &       & 0.6  & 0.35 & 0.3   & 0.2   & 
(9,7) & 0.0   & 0.0   & 0.0  & 0.0  & 0.2   & 0.0   \\
(6,6) & 0.05  & 0.05  &      &      & 0.3   & 0.2   & 
      &       &       &      &      &       &       \\
(7,1) &       &       & 0.4  & 0.1  &       &       &
      &       &       &      &      &       &       \\
(7,3) & 0.2   & 0.1   &      &      & 0.9   & 0.3   & 
      &       &       &      &      &       &       \\
(7,5) &       &       & 0.1  & 0.0   & 0.2  & 0.05  & 
      &       &       &      &       &      &       \\
(8,2) &       &       & 0.05  & 0.0  & 0.2   & 0.0   &
      &       &      & &&
\end{tabular}
\end{ruledtabular}
\end{table}

\clearpage
\begin{figure}
\caption{
Energy surfaces
of the critical IBM Hamiltonian, Eq.~(\ref{hcri}), 
with $\kappa=0.2$ and $N=10$. 
(a)~Intrinsic energy surface
$E(\beta)\equiv E(\beta,\gamma=0)$,
Eq.~(\ref{enesurf}), [solid line]. 
The unmixed $L=0$ and $L=2$ levels are shown for illustration. 
(b) $E(\beta)$ [solid line] as in~(a), 
unmixed $L=0$ [dashed line] and $L=2$ 
[long dashed line] projected energy surfaces, 
$E_{L}(\beta)\equiv E^{(N)}_{L}(\beta)$, 
Eq.~(\ref{egL}), and 
the lowest $L=0$ eigenpotential [dot-dashed line], 
$E^{(-)}_{L=0}(\beta)$, Eq.~(\ref{evL0}), whose global minimum is at 
$\beta=0.591$.}
\end{figure}


\newpage

\begin{thebibliography}{11}

\bibitem{iac00}
F. Iachello,
Phys. Rev. Lett. {\bf 85}, 3580 (2000).

\bibitem{iac01}
F. Iachello,
Phys. Rev. Lett. {\bf 87}, 052502 (2001).

\bibitem{voj03}
M. Vojta, Rep. Prog. Phys. {\bf 66}, 2069 (2003).

\bibitem{ibm}
F.~Iachello and A.~Arima,
{\it The Interacting Boson Model} 
(Cambridge Univ. Press, 1987).

\bibitem{vibron}
F.~Iachello and R.D.~Levine,
{\it Algebraic Theory of Molecules} 
(Oxford Univ. Press, 1994).


\bibitem{casten01}
R.F. Casten and N.V. Zamfir,
Phys. Rev. Lett. {\bf 87}, 052503 (2001).

\bibitem{krucken02}
R. Kr\"{u}cken {\it et al.}, 
Phys. Rev. Lett. {\bf 88}, 232501 (2002).

\bibitem{caprio02a}
M.A. Caprio, {\it et al.},
Phys. Rev. C {\bf 66}, 054310 (2002).

\bibitem{tonev04}
D. Tonev {\it et al.},
Phys. Rev. C {\bf 69}, 034334 (2004).


\bibitem{diep80}
A.E.L. Dieperink, O. Scholten and F. Iachello,
Phys. Rev. Lett. {\bf 44}, 1747 (1980).

\bibitem{izc98}
F. Iachello, N.V. Zamfir and R.F. Casten,
Phys. Rev. Lett. {\bf 81}, 1191 (1998).

\bibitem{ckz99}
R.F. Casten, D. Kusnezov and N.V. Zamfir,
Phys. Rev. Lett. {\bf 82}, 5000 (1999).

\bibitem{zam02}
N.V. Zamfir, P. von Brentano, R.F. Casten and J. Jolie, 
Phys. Rev. C {\bf 66}, 021304(R) (2002).

\bibitem{levgin03}
A. Leviatan and J.N. Ginocchio,
Phys. Rev. Lett. {\bf 90}, 212501 (2003).

\bibitem{arias03}
J.M. Arias, J. Dukelsky and J.E. Garc\'{\i}a-Ramos, 
Phys. Rev. Lett. {\bf 91}, 162502 (2003).

\bibitem{iaczam04}
F. Iachello and N.V. Zamfir,
Phys. Rev. Lett. {\bf 92}, 212501 (2004).

\bibitem{rowe04}
D.J. Rowe, P.S. Turner and G. Rosensteel,
Phys. Rev. Lett. {\bf 93}, 232502 (2004).

\bibitem{gino80}
J.N. Ginocchio and M.W. Kirson,
Phys. Rev. Lett. {\bf 44}, 1744 (1980). 

\bibitem{kirlev85}
M.W. Kirson and A. Leviatan,
Phys. Rev. Lett. {\bf 55}, 2846 (1985).

\bibitem{Hag03}
K. Hagino, G.F. Bertsch and P.-G. Reinhard, 
Phys. Rev. C {\bf 68}, 024306 (2003).

\bibitem{jolie99}
J. Jolie, P. Cejnar and J. Dobe\v{s}, 
Phys. Rev. C {\bf 60}, 061303(R) (1999).

\bibitem{cap99}
Jing-ye Zhang, M.A. Caprio, N.V. Zamfir and R.F. Casten,
Phys. Rev. C {\bf 60}, 061304(R) (1999).

\bibitem{frank04}
A. Frank, P. Van Isacker and C.E. Vargas,
Phys. Rev. C {\bf 69}, 034323 (2004).


\end{thebibliography}
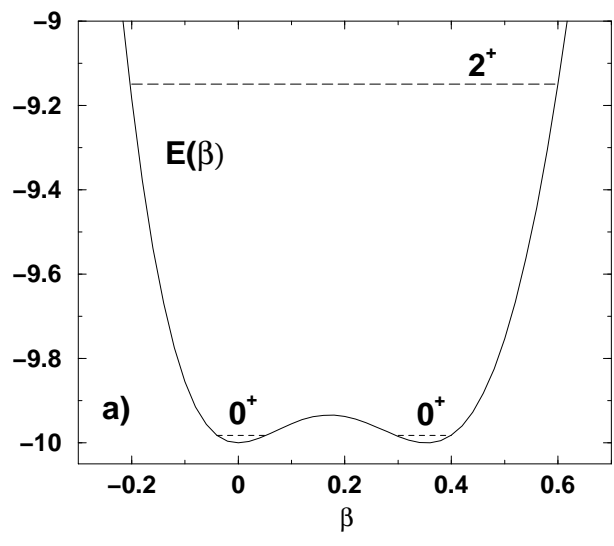
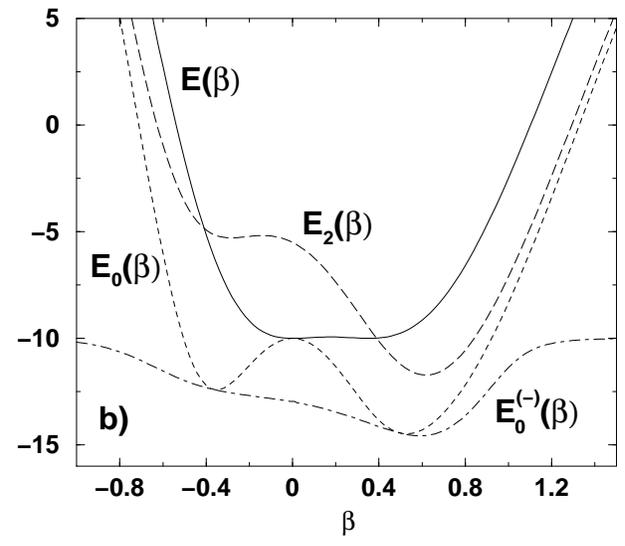
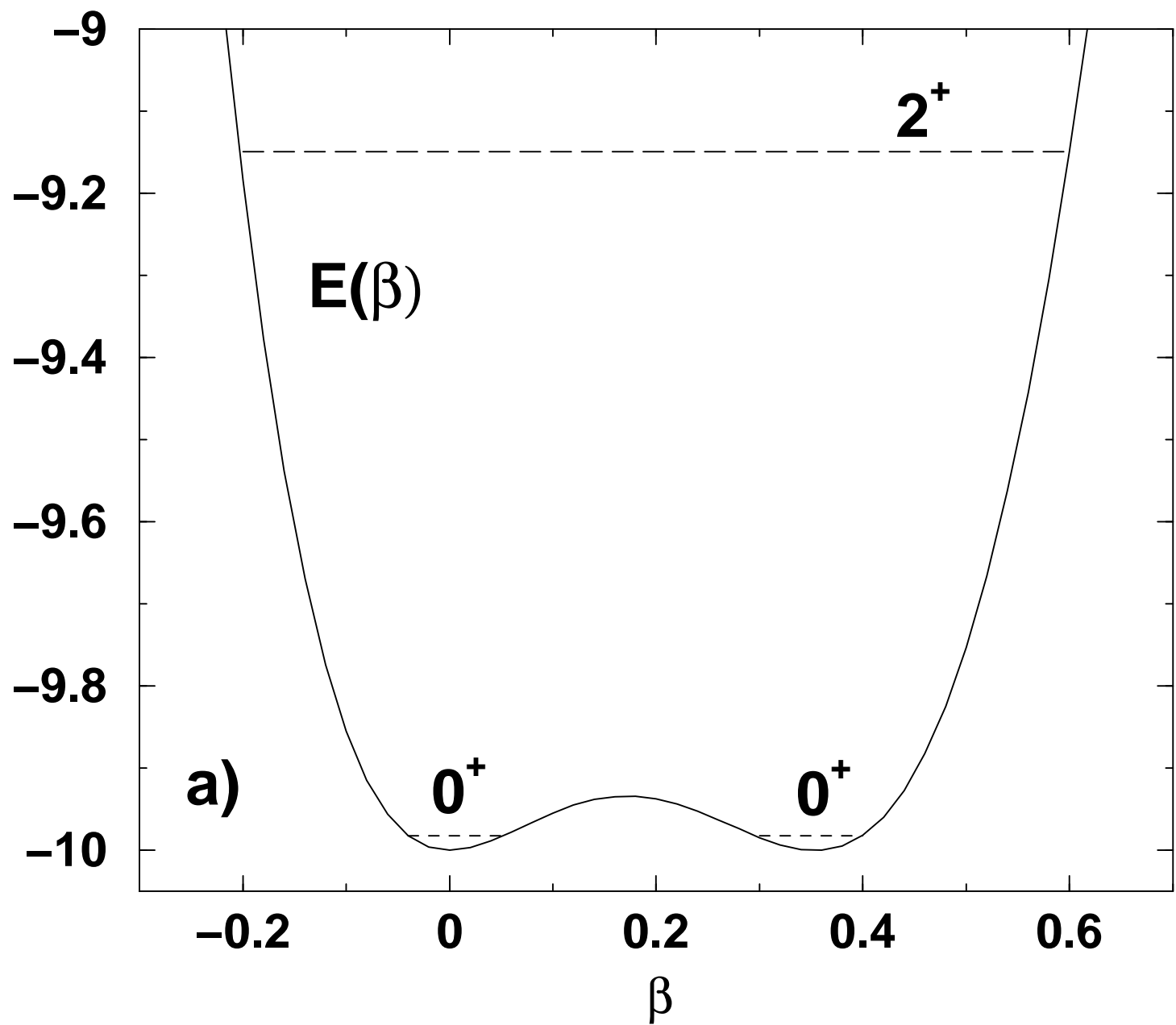
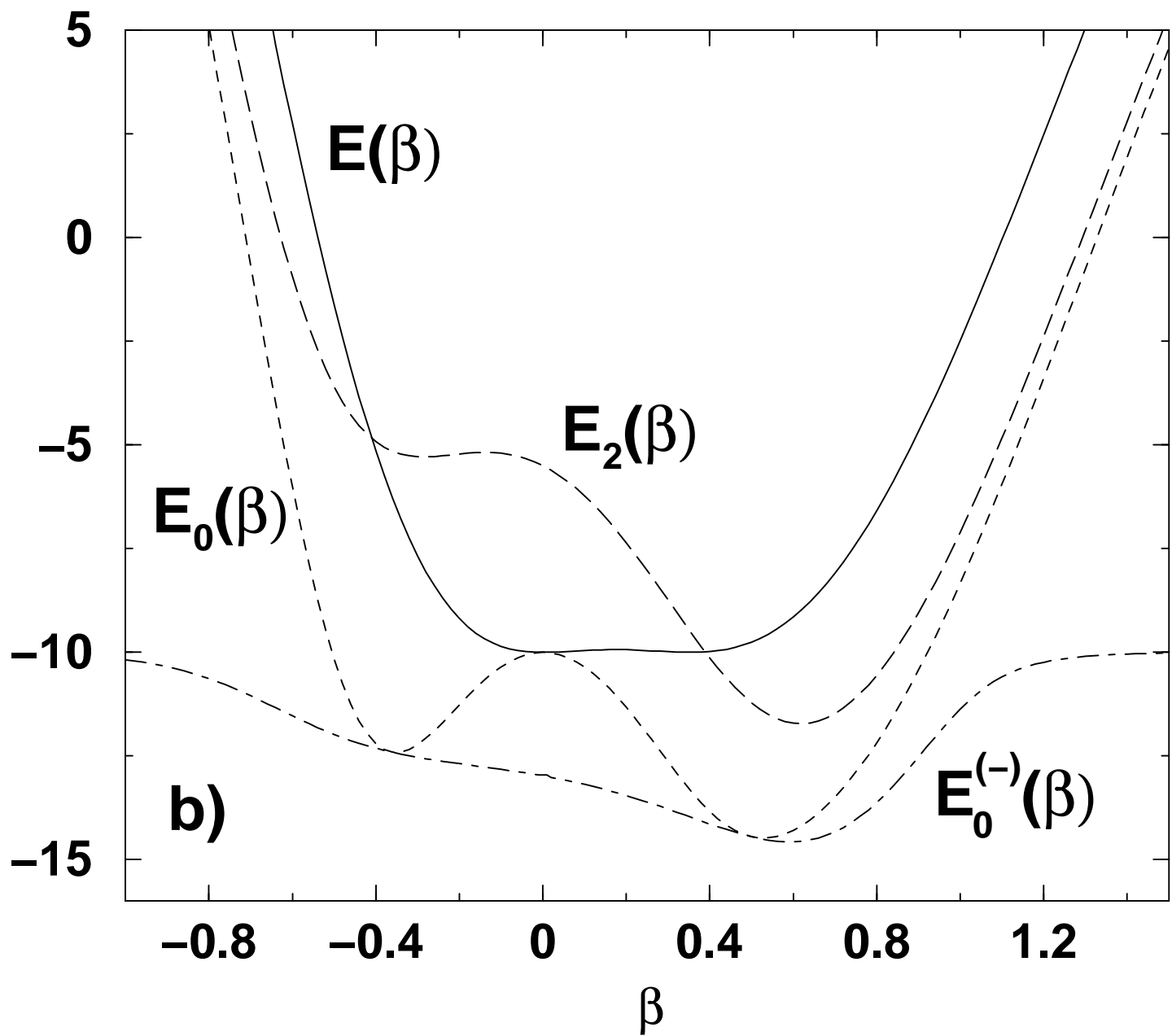
\end{document}